\documentclass[conference]{IEEEtran}
\IEEEoverridecommandlockouts

\usepackage{cite}
\usepackage{amsmath,amssymb,amsfonts}
\usepackage{graphicx}
\usepackage{textcomp}
\usepackage{xcolor}
\usepackage{url}
\usepackage{float}

\begin{document}

\title{Making Deployments Safe at Meta: \\
Health Checks for Continuous Change-Safety}

\author{%
\IEEEauthorblockN{Prakash KL}
\IEEEauthorblockA{\textit{Meta Platforms, Inc.}\\
pkl@meta.com}
\and
\IEEEauthorblockN{Anton Korenkov}
\IEEEauthorblockA{\textit{Meta Platforms, Inc.}\\
korenkov@meta.com}
\and
\IEEEauthorblockN{Uttam Thakore}
\IEEEauthorblockA{\textit{Meta Platforms, Inc.}\\
uthakore@meta.com}
\and
\IEEEauthorblockN{Christopher Hegre}
\IEEEauthorblockA{\textit{Meta Platforms, Inc.}\\
hegrec@meta.com}
}

\maketitle

\begin{abstract}
Continuous deployment to large-scale production systems creates a tension between release velocity and reliability: every change is a potential reliability incident, yet every delay is a missed opportunity. This paper describes the deployment-time health-check infrastructure that Meta uses to mediate this tension across thousands of heterogeneous services. We summarize the architecture of the Service Health Checker (SHC); explain how check authors compose templated metric queries, thresholds, and workflow predicates; and discuss how the system is integrated with tiered and phased rollouts so that regressions trigger automatic rollback. We then describe the operational problems that emerged at scale, such as noise, alert fatigue, drift, and uncovered regressions, and the program of measurement, tooling, and improved defaults we deployed to address them. We close with lessons learned from years of operating deployment health checks at Meta, and the directions we are exploring next, including AI-assisted health check tuning.

\end{abstract}

\begin{IEEEkeywords}
deployment safety, continuous deployment, monitoring, software reliability, release engineering, software reliability engineering, AIOps, anomaly detection
\end{IEEEkeywords}

\section{Introduction}
\label{sec:intro}

The reliability of modern internet services depends, in the most direct sense, on what happens during the few minutes a code or configuration change is being rolled into production. This is not a problem unique to any single organization: industry reports document that configuration and deployment changes are the single largest category of outage triggers across cloud providers~\cite{thousandeyes2024}. High-profile outages underscore the severity; Meta's October 2021 global outage, triggered when a routine backbone configuration change withdrew BGP routes, rendered all services unreachable for six hours~\cite{meta_outage_2021}, and the July 2024 CrowdStrike Falcon sensor update crashed 8.5 million Windows hosts worldwide, grounding airlines and disrupting hospitals across multiple continents~\cite{crowdstrike2024}. These incidents illustrate that even mature organizations with extensive testing infrastructure remain vulnerable when deployment-time validation is inadequate.

Pre-merge defenses (unit tests, integration tests, performance benchmarks) catch a large fraction of defects, but a non-trivial residue only manifests under production load, in production data shapes, against production dependencies. Recent work on deployment risk reinforces that pre-production testing alone is insufficient for systems operating at planetary scale. The last line of defense before a regression reaches users is therefore the rollout pipeline itself, and specifically the \textit{deployment-time health checks} that monitor the system while a change is being applied.

This paper provides an experience report about Service Health Checker (SHC), the system we have built at Meta to enable deployment-time health checking at scale, and the challenges we faced and addressed over several years of operating deployment health checks across thousands of internal services. 
Our intended audience is practitioners who run their own continuous-delivery infrastructure and researchers who study the engineering practice of large-scale software operations.

Three observations frame the paper. First, the design of a check is inseparable from the design of the rollout: a check that is meaningful at one phase of the rollout (e.g., a comparison against a control group) is meaningless at another (e.g., when no control group exists). Second, the dominant operational cost of health checks is not the regressions they catch, but the \emph{false positives} they raise: an unattended check that fires often is quickly demoted by its owners and ceases to protect them. Third, neither the system nor the metric it monitors is static: services evolve daily, so the bar for ``healthy'' must evolve with them. Throughout the rest of the paper, we expand on these observations and show how they shaped our architecture, our tooling design, and the overall reliability of our systems through safe rollouts.

The contributions of our paper are as follows:
\begin{enumerate}
\item a description of the query-templated, threshold-driven health-check system deployed within Meta (\S\ref{sec:arch}--\ref{sec:integration});
\item a candid account of the operational pathologies that emerged at scale and their root causes (\S\ref{sec:challenges});
\item a description of the quality improvement program that we instituted to mitigate them (\S\ref{sec:improvements}); and
\item the lessons we learned and open problems we are pursuing (\S\ref{sec:lessons} and \S\ref{sec:future}).
\end{enumerate}

We expect these contributions to be useful to practitioners who run their own continuous-delivery infrastructure and researchers who study the engineering practice of large-scale software operations.

\begin{figure*}[!t]
\centerline{\includegraphics[width=\textwidth]{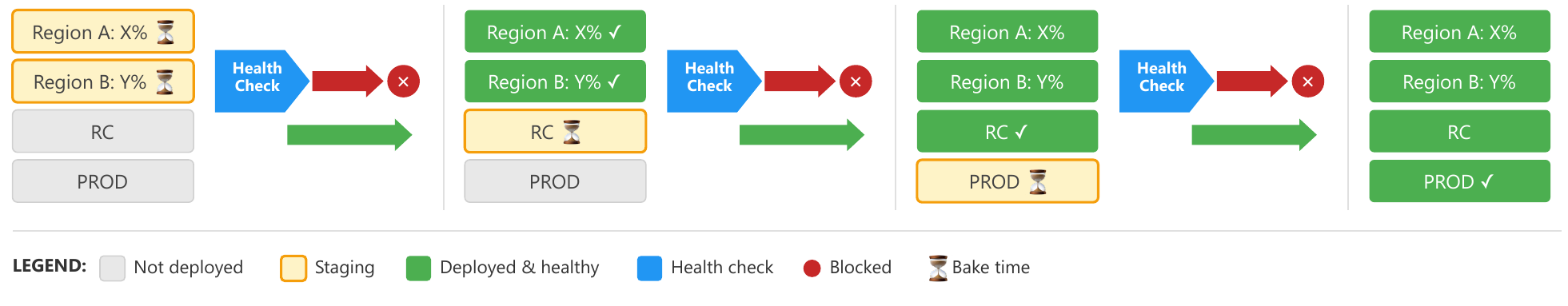}}
\caption{Progression of health checks throughout the deployment lifecycle. At each phase or tier of deployment, SHC monitors metrics for the hosts where changes have been rolled out for a predefined period (i.e., bake time) and runs health checks to confirm the health of the system post-deployment. On a failure verdict, the pipeline rolls back or pauses depending on the deployment strategy, and notifies the service owner.}
\label{fig:prod}
\end{figure*}

\section{Background: Deployment at Scale}
\label{sec:background}

Meta operates thousands of distinct services, libraries, and binaries, each deployed by one of several deployment systems specialized to its kind of artifact (e.g., binary push, configuration push, mobile client release). The services differ in topology, traffic mix, statefulness, and tolerance to disruption, so a single deployment strategy is not appropriate for all of them. What is common is the requirement that no rollout shall regress reliability.

At a high level, our continuous delivery systems employ two composable rollout strategies to enable safe deployments:

\subsubsection{Tiered rollouts}
Changes are first exercised in a pre-production release candidate (RC) tier that approximates production via traffic shadowing. Pre-production tiers cannot perfectly match production, as scale, hardware mix, and downstream load are difficult to reproduce, but they are inexpensive places to find obvious defects. In literature, this is also commonly referred to as ``staging'' or ``shadow testing''~\cite{savor2016continuous}.

\subsubsection{Phased production rollouts}
Once a change clears RC, it is rolled out to production in phases: a small number of hosts or a small region first, expanding progressively as confidence is gained. This bounds the blast radius of any single bad change and, equally important, gives the system observable time-windows in which a regression can be detected against the rest of the fleet. This is also commonly referred to as ``blue-green testing''.

In practice, both of these strategies are configurable: simple services may skip the tiered rollout strategy for the sake of speed, and more complex services may have multiple tiers and many phases within each tier~\cite{rossi2017rapid,petrochko2021conveyor}. Fig.~\ref{fig:prod} illustrates a deployment that has two regional canary phases, an RC tier, and a production tier.

The critical question, in both regimes, is the one our health checks are designed to answer at each tier/phase boundary: \emph{is the change exhibiting any signal of degradation, and if so, what should the deployment system do about it?} Answering this is harder than it appears. Some regressions are loud (crashes, error spikes); others are slow burns visible only after hours of cumulative traffic. The deployment cannot wait indefinitely: other deployments are queued behind it, the originating engineer is waiting, and other systems are themselves in flux. A correct decision must therefore be made under a time budget, with imperfect signal, and with the rest of the production fleet acting as a noisy baseline.

\section{Health Check Architecture}
\label{sec:arch}

A health check, in our system, is a configuration object that binds a metric query to a pass/fail rule and a workflow predicate. Every service is bootstrapped with a default set of checks covering generic indicators (CPU and memory utilization, crash counts, categorized error ratios) so that even a team that writes no custom checks has some safety net. Service owners then add on application-specific checks that reflect the things their service is supposed to do, such as per-dependency latencies, request cost, and business-level error budgets.

\begin{figure*}[tb]
\centerline{\includegraphics[width=\textwidth]{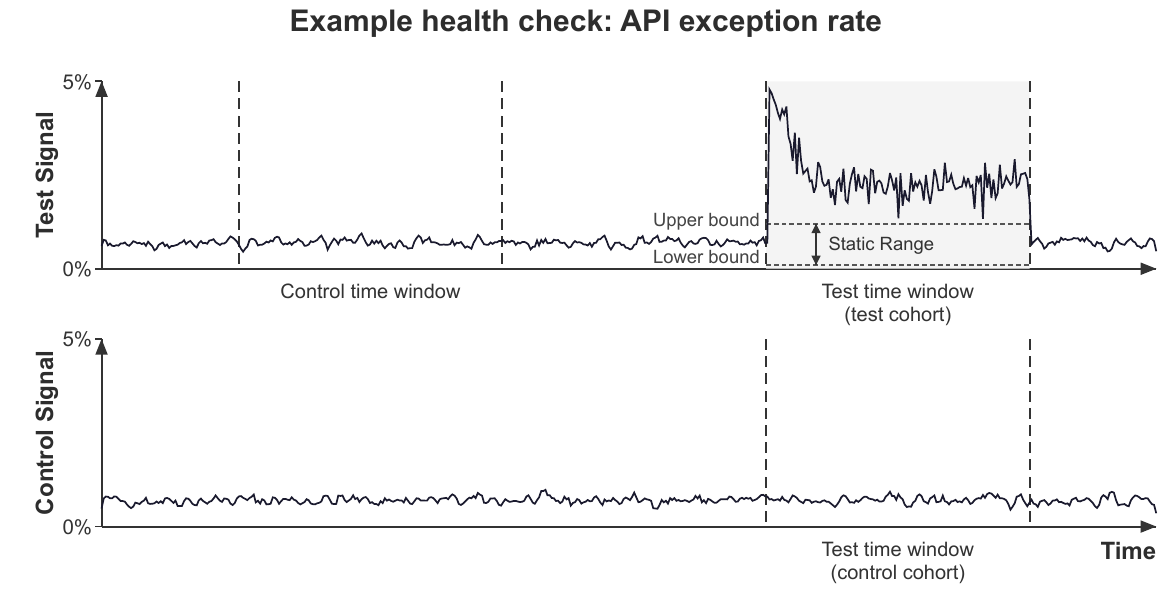}}
\caption{Anatomy of a health check. A templated metric query yields a time series for each of the test and control cohorts. The shaded region represents the time when the change is being deployed. The health check can define both \textit{static} bounds, outside of which the series is considered unhealthy, and \textit{comparison} thresholds, which compare the test signal during the test time window against either the test signal during a control time window (in the case of single-host services) or the control signal during the test window (in the case of multi-host services).}
\label{fig:check}
\end{figure*}

A check has three components (Fig.~\ref{fig:check}):

\subsubsection{Templated metric query}
The metric is expressed in a query language that returns timestamped data points. Crucially, the query is \emph{templated} on dimensions populated by the deployment system at evaluation time: the set of hosts that received the change, the set of hosts in the matched control population, the rollout identifier, and so on. This indirection lets a service owner write a single query (``request error rate'') without knowing whether, on a given run, it will be evaluated against a 1\% canary or a regional cohort.

\subsubsection{Thresholds}
Thresholds turn time series into pass/fail verdicts. The simplest are \emph{static} bounds (``error rate must be below $x$''), useful for catching gross regressions on stable signals. More useful in practice are \emph{comparison} thresholds, which compare the cohort under test to a control cohort drawn from the unchanged fleet. Comparison thresholds dominate static thresholds for slow-burn and traffic-dependent regressions because they automatically absorb diurnal and weekly seasonality and traffic spikes that would either trigger spurious alerts or require painstakingly tuned static bounds. A threshold may also specify a tolerated fraction of out-of-bound data points, and a policy for the absent-data case.

\subsubsection{Workflow predicate}
A check declares which deployment phases it applies to. Some checks make sense only in production (RC traffic is too sparse for the metric to be meaningful); others should be skipped on the smallest canary slices, where statistical power is too low to draw a conclusion. The predicate gives the check author explicit control over false positives that arise from running a check in the wrong deployment phase.

These three pieces compose into the health check definition that a service owner actually edits and version-controls. From the owner's perspective, the deployment system is, in effect, a consumer of these definitions.

To illustrate, take a hypothetical CPU-bound microservice where the owner would want to guard against changes that introduce the risk of overload or elevated request failures. The service might then have 3 health checks, one each for host-level CPU utilization, memory utilization, and API exception rate. The upper and lower bounds for the threshold for CPU and memory utilization might be 0-90\% to guard against throttling and out-of-memory errors, and for API exception rate to be 0-1\% (if the service provided a two-nines SLA). This check would be run in every deployment phase.

\section{Integration with Deployment Pipelines}
\label{sec:integration}

The component that evaluates checks is the Service Health Checker (SHC). When a deployment pipeline transitions between phases, it issues a request to SHC identifying the cohort under test and the rollout context. SHC fetches the relevant metrics, applies the configured thresholds, and returns a structured verdict that the pipeline acts on (Fig.~\ref{fig:prod}).

The action taken on a failure verdict is policy-defined per deployment system: most production pipelines roll back to the previous version on failure, while some pause and escalate to the on-call engineer. In either case, the service owner is notified with the failing check, the offending metric, and a link to the dashboard that breaks the rollout down by phase and cohort. Templating lets SHC isolate the affected slice precisely, which is what makes the rollback cheap: the system rolls back \emph{only} what was rolling forward.

SHC monitors and prevents bad rollouts on the full breadth of Meta's production systems: the web tier (www), messaging services, user-facing APIs serving sub-100ms latency budgets, machine learning inference services, storage backends, and internal developer tooling, each with distinct traffic patterns, failure signatures, and health indicators.

Every code and configuration change entering Meta production passes through SHC as a mandatory gate. The system prevents a significant number of high-confidence production incidents annually, specifically regressions that passed all pre-merge testing and would only otherwise have been caught during production rollout, representing substantial avoided revenue loss and operational cost. SHC is also cheap to operate: each host used by SHC is able to monitor thousands of production hosts.

A subtle but important property of this design is that check authors do not need to understand the deployment topology. The owner of a service writes a check against the metrics they care about; SHC, in concert with the deployment system, decides which dimensions to bind and how to draw the control cohort. This separation of concerns has been central to keeping the system tractable for non-specialist owners.

\section{Operational Challenges}
\label{sec:challenges}

The architecture above answers ``how do we run a check at the right time?'' What it does not answer is ``how do we know the checks are any good?'' Once SHC was widely adopted, two classes of problem dominated the operational experience.

\subsubsection{Noise}
A check that fires often without indicating a real problem is worse than no check at all, as it trains its owner to ignore it. In our experience, noise had several recurring causes. Thresholds were sometimes set so tight that ordinary fluctuation would breach them. Some checks ran on cohorts too small for the underlying metric to have meaningful resolution, producing high-variance results that crossed thresholds at random. Others were sensitive to flakiness in upstream or downstream dependencies, raising alarms that were structurally outside the owning team's ability to fix.

Noise had a number of operational consequences. First, noise caused alert fatigue: when checks fired frequently and most were false, engineers stopped investigating. Second, noise reduced velocity: every false positive paused or rolled back a deployment, lengthening the queue behind it and pushing more partially-deployed states into the fleet. Third, noise reduced carrying capacity for on-call engineers: investigating false alarms adds to on-call load, and on-call work that does not produce learning is the kind of work that drives engineers to file paperwork to delete the check.

\subsubsection{Drift}
A check tuned correctly today may be miscalibrated tomorrow because the service it protects is changing daily: workload composition shifts, new dependencies are added, an underlying library begins emitting a new error class. The owners' focus moves elsewhere, even as the check stays as it was. Over time, the population of deployed checks accumulates a long tail of items that no longer measure what they were written to measure.

The temptation, given the visibility of noise, is to tune for it. Doing so naively trades one failure mode for another: a check loose enough never to fire is not catching anything either. Recall and precision are both quality dimensions that trade off against each other. We needed a way to measure and steer that trade-off rather than just reacting to whichever was complaining loudest.

\section{Quality Improvement}
\label{sec:improvements}

Our response to the operational challenges above had two dimensions: human-in-the-loop quality improvement and automated quality improvement.

\begin{figure}[!t]
\centerline{\includegraphics[width=\columnwidth]{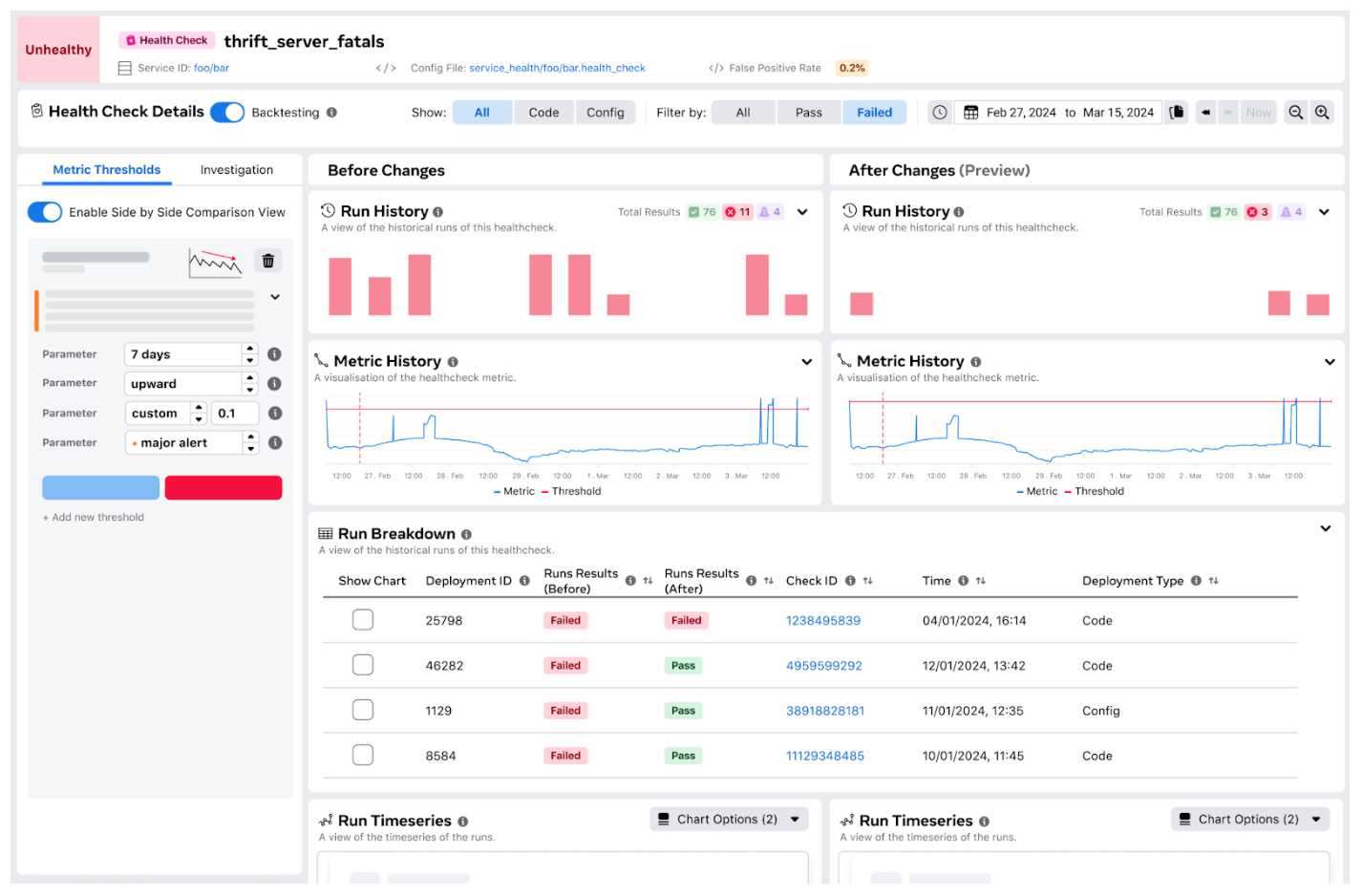}}
\caption{Backtesting tool that enables service owners to test a candidate check against historical data. The owner can see how the proposed definition would have classified past incidents and past noise before the check is deployed.}
\label{fig:backtest}
\end{figure}

\subsection{Human-in-the-Loop Quality Improvement}

We built a daily data pipeline 
that joins three streams: the verdicts SHC has issued, the deployment outcomes that followed those verdicts, and a corpus of explicit and implicit ground-truth signals (e.g., manual disposition of an alert, whether a rollback was followed by a re-attempt of the same change). From these inputs the pipeline categorizes each verdict as a true or false positive, and rolls the labels up into per-check precision and recall estimates.


Each check is then surfaced to its owner in a catalog view that flags low-quality checks at a glance, with the false-positive rate as the primary indicator and a longer list of secondary best-practice violations (e.g., absent control cohort, missing absent-data policy). This enables owners to quickly act on low quality health checks without needing to invest in data analysis.

Knowing a check is bad is not enough; the owner has to be able to fix it without disrupting the service it protects. We therefore built a backtesting workflow (Fig.~\ref{fig:backtest}) in which a candidate check definition is replayed against 30 days of full-fidelity historical metric data (retained at 1-minute granularity) covering both ordinary days and known incidents. Before deploying the change, the owner can see how the new definition would have behaved on past noise and on past real regressions. This makes the precision/recall trade-off visible at edit time, which is the only point at which the owner is in a position to act on it.

Beyond per-check tooling, we evolved the bootstrapped defaults each new service starts with, so that the population of checks improves even for owners who never touch them. We also invested in non-tooling mechanisms, like an internal community for check authors to share patterns, high quality documentation, and training. These low-tech mechanisms turned out to matter as much as the dashboards: noise is in part a sociological problem, and the fix must have a sociological component.

\subsection{Automated Quality Improvement}

In addition to empowering service owners to improve health checks manually, we also invested in automated improvements that did not require service owner participation.

\subsubsection{Topology-aware metric publication}
\begin{figure}[tb]
\centerline{\includegraphics[width=1.05\columnwidth]{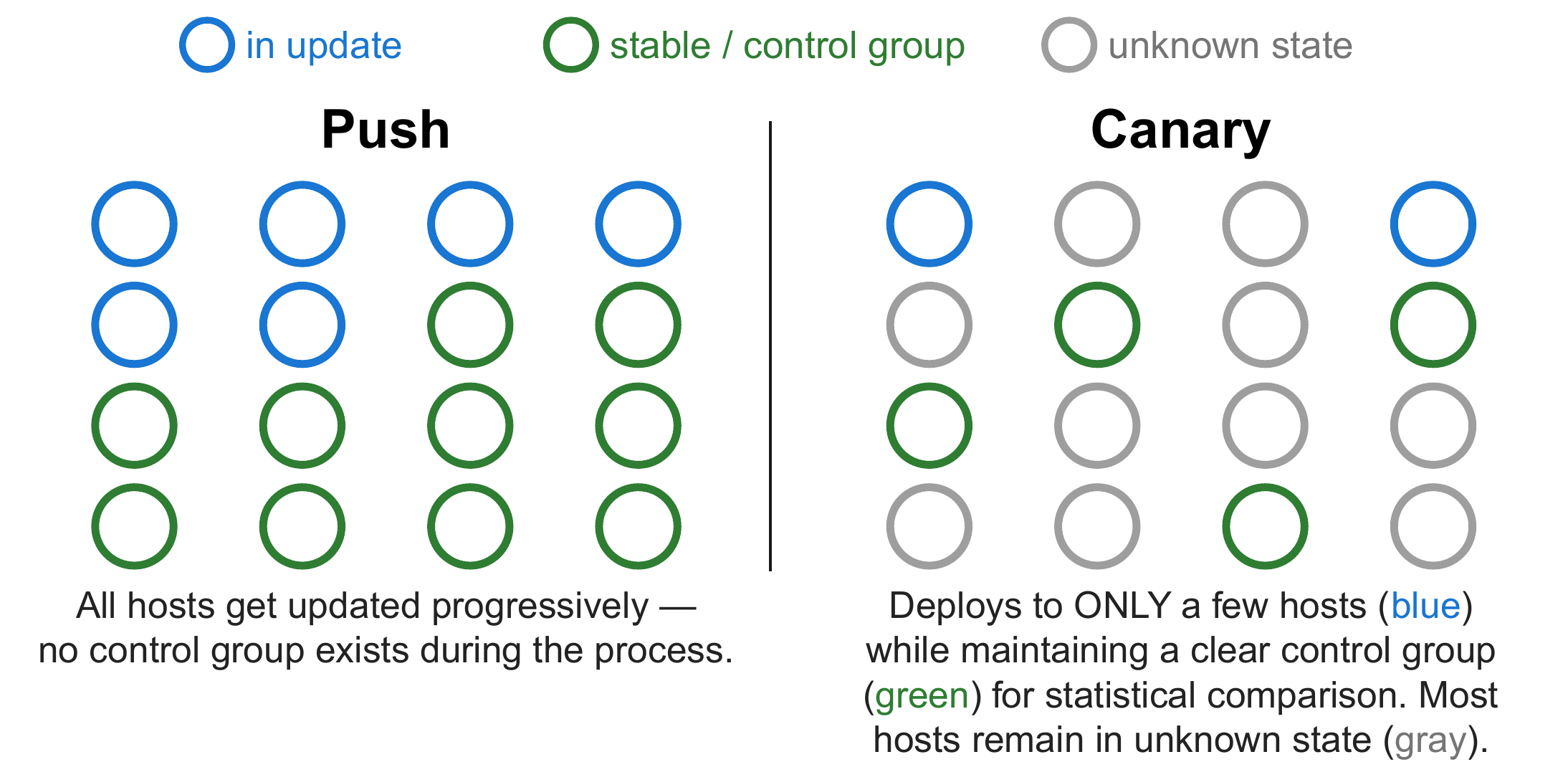}}
\caption{Metric targeting. Templated cohorts let SHC evaluate the same health check at the right granularity for each deployment phase, from a small canary slice up to a regional cohort.}
\label{fig:target}
\end{figure}

We rationalized the dimensions on which metrics are published so that the same metric can be evaluated against the right cohort at every phase of every deployment system, without each service redefining its own slicing. Accurate targeting (Fig.~\ref{fig:target}) is the single largest lever for detection latency: the sooner a regression appears in a slice that is small enough to roll back cheaply, the lower the cost of catching it.

\subsubsection{Severity-aware verdicts}
A check that fails by a hair on a small slice is not the same risk as one that fails by an order of magnitude on a regional cohort, and the deployment system should have the option to react differently. We are extending verdict semantics to expose how badly a threshold was breached and how widely the breach was observed, so that downstream policies can respond proportionally rather than treating every failure as equivalent.

\subsubsection{Dependency-aware checking}
A change to one service can degrade upstream callers or downstream dependencies that the deploying owner does not own and may not be aware of. We mine an offline dataset of cross-service correlations to identify each service's critical dependencies, and SHC automatically evaluates checks belonging to those dependencies in addition to the deploying service's own checks. This catches a class of cross-service regressions that would otherwise be invisible to the local check set.

\subsubsection{SLI-based gating}
Symptoms are not always a sufficient leading indicator; sometimes the only reliable signal is the service's SLI itself. We provide a uniform mechanism for any service to declare its SLIs and have SHC gate deployments on them, so that an SLI regression is enough on its own to halt a rollout.

\paragraph*{Quantitative Impact}



Over the course of one year, the quality improvement program reduced the fleet-wide false positive rate from 12.1\% to 2.7\%. An independent validation (two senior engineers classifying 500 random failures) confirmed the self-reported rate with a modest 0.5\% underreporting bias.

\section{Lessons Learned}
\label{sec:lessons}

\textbf{Default behavior is rarely modified.}
Most service owners never touched the bootstrapped default check, so it became the most common check in production. Investments that improved the defaults dominated, in aggregate, investments that helped any individual owner write a better custom check. We underweighted this for too long.

\textbf{Quality requires measurement, not just tooling.}
The transition from anecdotal quality to hard numbers (precision and recall, surfaced per check) was necessary to make conversations about health check quality tractable across the company. Prior to this, teams would claim their checks were good with little evidence to back their claims.

\textbf{Health check quality needs to be presented as a trade-off.}
Service owners want a single answer (``is this check good?''), but the honest answer that health checks lie on a precision-recall curve along which they have to choose. Backtesting was effective because it presented the trade-off in concrete terms (this many past incidents caught, this many past false alarms produced) rather than as a pair of abstract metrics.

\textbf{Metric templating makes health check authorship easier.}
Letting a check author write a query without knowing how it would be sliced for evaluation was, in retrospect, the single most consequential decision in the architecture. It made it feasible to run the same check across heterogeneous deployment systems without forcing owners to understand the nuances of each one.

\textbf{Service dependency introduces fragility.}
Service owners can write checks against their own metrics, but the regressions that hurt most often cross service boundaries. Dependency-aware checking is important to catch such regressions because health checks on upstream service rollouts may mask regressions on specific request paths that are only exercised by downstream services.

\textbf{Sociotechnical investments pay off.}
Comprehensive documentation, a chat channel, and an office-hour rotation moved more checks from low quality to acceptable quality than several quarters of dashboard work, because they targeted the part of the problem (owners not knowing what good looks like) that no dashboard could fix on its own.

\section{Related Work}
\label{sec:related}

Continuous delivery as a practice is canonically articulated by Humble and Farley~\cite{humble2010cd}. Adams and McIntosh~\cite{adams2016saner} survey academic literature on continuous delivery and argue that release engineering is an under-studied area of software-engineering research. Schermann et al.~\cite{schermann2018ist} conduct a multi-method empirical study of continuous experimentation in industry, including how teams gate rollouts with metrics. Our work is in the same lineage but focuses specifically on the \emph{evaluation substrate} that decides whether a phase of a rollout proceeds.

The most direct point of comparison for SHC is Google's Canary Analysis Service, described by Davidovic and Beyer in \cite{davidovic2018canary}. The two systems share the design pattern of comparing a cohort under test against a control cohort using statistical thresholds; SHC additionally exposes templated workflow predicates that select which checks apply at which deployment phase, and ties verdicts back to per-check precision and recall surfaced in an owner-facing catalog. Google's SRE handbook~\cite{beyer2016sre} codifies the SLO-centric reliability discipline that motivates our SLI-based gating. Tarvo et al.~\cite{tarvo2015canaryadvisor} present an early statistical canary tool whose threshold-tuning workflow anticipates several of the same trade-offs we discuss.

We build on prior work at Meta on production safety at scale. Tang et al.~\cite{tang2015holistic} describe Meta's earlier work on configuration deployment, and Veeraraghavan et al.~\cite{veeraraghavan2016kraken,veeraraghavan2018maelstrom} describe complementary mechanisms (live traffic testing, datacenter drains) that operate at coarser granularities than SHC.

There is also extensive literature on how to detect and localize regressions from production telemetry~\cite{xu2009console,sambasivan2011diagnosing,mace2015pivot,liu2019fluxrank}, and recent surveys catalog the broader AIOps space~\cite{notaro2021aiops}. SHC enables service owners to leverage these techniques. Our experience with the precision/recall trade-off (\S\ref{sec:improvements}) is consistent with that literature and motivates our move toward learned thresholds in future work.

\section{Future Work}
\label{sec:future}

Our most active direction is the application of machine learning to improving check quality. We are attempting to learn threshold and metric selections from the historical signal we already log for the efficacy pipeline, and adapt them as the underlying service drifts. We are piloting a gradient-boosted model on approximately 5\% of services that predicts per-deployment failure probability from metric features (recent check history, metric variance, deployment size, service archetype). Early results show a 30\% reduction in recommended threshold tightness while maintaining recall, though the pilot has not yet been deployed to the owner-facing workflow. The same data (verdicts, deployment outcomes, and ground-truth labels) that power measurement also power training, so we can move incrementally from measurement-driven manual tuning toward measurement-driven \textit{automatic} tuning, with humans in the loop where the cost of an error is high. Our long-term aim is a self-tuning check population in which an owner's intervention is required only at service inception or after a structural change.

In parallel, we are exploring using distributed tracing as a complementary signal to the metrics SHC uses today. Traces give us causal attribution that aggregate metrics do not, which we expect to shorten investigation time when a check does fire and to surface dependency regressions earlier than the offline correlation analysis can.

\section{Conclusion}
\label{sec:conclusion}

Deployment-time health checks are the practical mechanism by which Meta has held the line on reliability while continuing to ship at high cadence. In this paper, we describe Service Health Checker, Meta's highly scalable system for deployment-time health checking, which is integrated directly with our continuous delivery systems. While SHC is conceptually simple, we have tackled various hard operational problems to make health checking effective and practical: keeping checks honest as services drift, distinguishing real regressions from noise without losing recall, and giving thousands of service owners the data and tooling to act on the difference. Our measurement-and-iteration programs have enabled service owners at Meta to meaningfully improve their health check quality and reduce operational overhead.

\section*{Acknowledgments}
Service Health Checker was built and refined by many engineers over the course of many years; this paper describes systems that are the cumulative product of their work. The authors would like to specifically thank Ruben Badaro, Alex Kindyakov, Ahmad Mamdouh Abdou, Ahmed Eid and the many other engineers who have contributed to Service Health for their contributions to the systems described in this paper.

\bibliographystyle{IEEEtran}
\bibliography{paper}

\end{document}